\documentclass[a4paper,twocolumn,unpublished]{quantumarticle}
\pdfoutput=1
\usepackage[numbers]{natbib}
\usepackage[dvips]{graphicx}
\usepackage{amsmath,amssymb,amsthm,mathrsfs,amsfonts,dsfont}
\usepackage{epsfig}
\usepackage{physics}
\usepackage{braket}
\usepackage{hyperref}

\hypersetup{
    breaklinks=true,   
    colorlinks=true,   
    pdfusetitle=true,  
 }
\usepackage{bm}
\usepackage{enumerate}
\usepackage{color}
\usepackage{graphicx}
\usepackage{pgf}
\usepackage{tikz}
\usepackage{yquant}
\usetikzlibrary{calc}
\usepackage{enumitem}
\usepackage[capitalize]{cleveref}
\usepackage[normalem]{ulem}
\usepackage{float}

\usepackage{xcolor}

\usepackage{amsthm}

\Crefname{theorem}{Theorem}{Theorems}

\usepackage{orcidlink}

\begin{document}


\title{Low-Overhead Parallelisation of LCU via Commuting Operators}

\author{Gregory Boyd \orcidlink{0000-0001-7822-1688}}
\affiliation{Department of Materials, University of Oxford, Parks Road, Oxford OX1 3PH, United Kingdom}


\begin{abstract}
The Linear Combination of Unitaries (LCU) method is a powerful scheme for the block encoding of operators but suffers from high overheads. In this work, we discuss the parallelisation of LCU and in particular the \textsc{SELECT} subroutine of LCU based on partitioning of observables into groups of commuting operators, as well as the use of adaptive circuits and teleportation that allow us to perform required Clifford circuits in constant depth. We additionally discuss the parallelisation of QROM circuits which are a special case of our main results, and provide methods to parallelise the action of multi-controlled gates on the control register. We only require an $O(\log n)$ factor increase in the number of qubits in order to produce a significant depth reduction, with prior work suggesting that for molecular Hamiltonians, the depth saving is $O(n)$, and numerics indicating depth savings of a factor approximately $n/2$. The implications of our method in the fault-tolerant setting are also considered, noting that parallelisation reduces the $T$-depth by the same factor as the logical algorithm, without changing the $T$-count, and that our method can significantly reduce the overall space-time volume of the computation, even when including the increased number of $T$ factories required by parallelisation.
\end{abstract}

\maketitle

\section{Introduction} \label{sec:intro}

The Linear Combination of Unitaries (LCU) method \cite{childsHamiltonianSimulationUsing2012} is a powerful quantum subroutine used for the efficient block encoding of operators, that when combined with Quantum Singular Value Transformation (QSVT) \cite{gilyenQuantumSingularValue2019, martynGrandUnificationQuantum2021a} leads to asymptotically optimal algorithms for Hamiltonian simulation \cite{lowHamiltonianSimulationQubitization2019a} as well as dissipative ground state finding algorithms \cite{chenEfficientExactNoncommutative2023}, and quantum linear algebra \cite{childsQuantumAlgorithmSystems2017}. Despite the power of LCU, there are high overheads associated with its implementation, leading to recent work promoting the use of product formula methods for Hamiltonian simulation as a practical alternative \cite{rendonImprovedErrorScaling2022,zengSimpleHighprecisionHamiltonian2022}. In this work, we make a contribution to reducing the time overhead of LCU by presenting a general method for parallelising the algorithm for operators provided in the Pauli representation that we conjecture  is typically able to reduce the depth ($T$-depth in the fault-tolerant setting) of LCU by a factor of $O(n)$ (see \cref{sec:partitioning}) while only requiring an $O(\log n)$ factor increase in qubit count, resulting in a net reduction in the space-time volume cost of the logical algorithm by a factor of $O(n/\log n)$. A special case of the circuits we propose parallelisation methods for in this work is QROM circuits \cite{babbushEncodingElectronicSpectra2018}, used for loading classical data into quantum computers.

We also provide methods for parallelisation of the application of gates with complicated patterns of positive and negative controls from an ancilla register of size $a$ at the cost of depth $O(\log a)$, which may be preferable when the size of the ancilla register is large (as opposed to the case assumed for LCU above where the register has size $O(\log n)$) which allows us to parallelise the above circuits with a constant factor qubit overhead, reducing the depth by a factor of $O(n/\log a)$.

Special cases of reducing the complexity of LCU exist, e.g. for the specific case of local fermionic Hamiltonians where tensor hypercontraction methods \cite{leeEvenMoreEfficient2021} and others \cite{wanExponentiallyFasterImplementations2021,vonburgQuantumComputingEnhanced2021a,berryQubitizationArbitraryBasis2019} can be used to reduce the daunting $O(n^4)$ cost of block encoding molecular Hamiltonians. Parallelisation methods also exist for performing Hamiltonian simulation for certain kinds of Hamiltonian \cite{zhangParallelQuantumAlgorithm2023a}.

Prior work also includes methods using one-hot or $k$-hot encodings of the coefficients or fan-out have been examined to reduce the depth and improve the errors in LCU by taking advantage of the parallelisability of geometrically local observables \cite{zeytinogluErrorRobustQuantumSignal2022,yoshiokaHuntingQuantumclassicalCrossover2023a}.
In \cref{sec:background} we discuss the LCU method, as well as the basis for the parallelisation techniques we will use. \cref{sec:parallelisation} will discuss our method in more detail, including discussion of the implementation of these methods on fault-tolerant hardware. 

\section{Background} \label{sec:background}

\subsection{LCU}

The LCU method constructs a linear combination of unitaries using the combination of \textsc{PREPARE} and \textsc{SELECT} subroutines. It involves constructing an operator $A$ as a linear combination of a set of unitary operators $U_j$ with corresponding coefficients $c_j$:

\begin{equation} \label{eq:LCU}
A = \sum_j^L c_j U_j
\end{equation}

This is done by loading the coefficients into an ancilla register in the \textsc{PREPARE} step:

\begin{equation} \label{eq:prepare}
    \ket{0^a} \rightarrow \ket{\alpha} = \frac{1}{\sqrt{\lVert c \rVert_1}} \sum_j \sqrt{c_j} \ket{j}
\end{equation}

where the size of the ancilla register is $\log L$. We will restrict our analysis when discussing LCU to $L\sim \text{poly}(n)$ so that the number of ancilla in the coefficient register is $O(\log n)$. \\
Then the unitaries are applied controlled on the ancilla in the \textsc{SELECT} step:

\begin{equation} \label{eq:SELECT}
    \ket{j} \ket{\psi} \rightarrow \ket{j} U_j \ket{\psi}
\end{equation}

\begin{figure}[H]
    \center
    \begin{tikzpicture}
    \begin{yquant}
        qubit {$\ket{j_\idx}$} a[2];
        qubit {$\ket{\psi}$} psi;
        slash psi;
        hspace {2mm} -;
        box {$U_0$} psi ~ a[-];
        box {$U_1$} psi | a[0] ~ a[1];
        box {$U_2$} psi | a[1] ~ a[0];
        box {$U_3$} psi | a[-];
    \end{yquant}
    \end{tikzpicture}
\caption{Example of the usual procedure for the \textsc{SELECT} step of LCU with two ancillas.}
\label{fig:SELECT}
\end{figure}
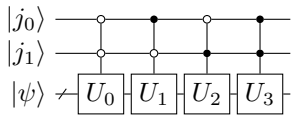

LCU is of particular use for implementing block encodings of non-unitary operators, such as Hamiltonians, and therefore provides access to performing QSVT on these block encodings to apply functions of the operator.

\subsection{QROM}

Quantum Read-Only Memory \cite{babbushEncodingElectronicSpectra2018} is a technique for loading classical data into quantum memory, by allowing data to be accessed by an index in superposition:

\begin{equation} \label{eq:qrom_definition}
    \text{QROM}_d \sum^{L-1}_{l=0} \alpha \ket{l}\ket{0} = \sum^{L-1}_{l=0} \alpha \ket{l}\ket{d_l}
\end{equation}

This can be achieved using a circuit in the form of \cref{fig:SELECT} where the unitaries applied are tensor products of Pauli $X$'s encoding the binary strings of the data entries $d_l$ to be loaded. 
QROM circuits are also used as a component of advanced schemes for block encoding quantum chemistry Hamiltonians \cite{leeEvenMoreEfficient2021}, where it is noted that QROM is the dominant cost of the procedure, and therefore a prime target for runtime optimisation.

\subsection{Parallelisation Techniques} \label{sec:parallelisation_techniques}

Techniques for parallelising quantum algorithms have been known for some time \cite{mooreParallelQuantumComputation1998}. Our algorithm makes use of the primitives of fanout and gate teleportation using states prepared by measurement and feedforward.

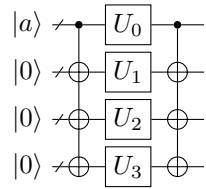
\begin{figure}[H]
    \center
    \begin{tikzpicture} 
        \begin{yquant}
            qubit {$\ket{a}$} a;
            qubit {$\ket{0}$} q[3];
            slash a;
            slash q;
            cnot q[0-] | a;
            box {$U_0$} a;
            box {$U_1$} q[0];
            box {$U_2$} q[1];
            box {$U_3$} q[2];
            cnot q[0-] | a;
        \end{yquant}
    \end{tikzpicture}
    \caption{Fanout operation performing operations on the same input in parallel.}
    \label{fig:fanout}
\end{figure}

The fanout operation \cite{hoyerQuantumCircuitsUnbounded2005} (\cref{fig:fanout}) can be thought of as a basis-dependent ``copy'' operation that takes information in the computational basis from one register and spreads it across additional registers. 

\begin{equation} \label{eq:fanout}
    \ket{j} \ket{0} \dots \ket{0} \rightarrow \ket{j} \ket{j} \dots \ket{j}
\end{equation}

Crucially, this allows us to perform commuting operations on a qubit register in parallel, by first performing a fanout to spread the information across multiple registers, and then performing the operations on each register individually. We then uncompute the fanout operation.

A log-depth fanout circuit can be constructed straightforwardly from a tree of CNOT gates, and a constant depth circuit can be constructed using teleportation with $O(k)$ ancillas, where the fanout acts on $k$ qubits \cite{buhrmanStatePreparationShallow2023a}.
\\

An additional tool we use is the use of measurement and feedforward (or adaptive circuits), where circuits contain measurements and operations that are controlled on the measurement results, resulting in deterministic measurement based circuits to prepare certain resource states \cite{buhrmanStatePreparationShallow2023a,foss-feigExperimentalDemonstrationAdvantage2023} and the use of gate teleportation with these states \cite{gottesmanQuantumTeleportationUniversal1999} to reduce circuit depth.

\subsection{Partitioning into commuting groups} \label{sec:partitioning}

A third technique used in our work is the partitioning of operators into commuting groups of Pauli terms, these methods were initially developed for the optimisation of measurements in variational algorithms \cite{crawfordEfficientQuantumMeasurement2021a, gokhaleMinimizingStatePreparations2019a, verteletskyiMeasurementOptimizationVariational2020, yenMeasuringAllCompatible2020}. In this context, it is desirable to partition observables into commuting groups to determine which components can be measured simultaneously. This is done by calculating an approximate Minimum Clique Cover (MCC) of the graph $G=(V,E)$ where the vertices are the Pauli terms in the operator and edges are placed between commuting terms. \\

The amount of commuting terms found using this method will of course depend on the operator in question, but evidence from numerics on molecular electronic Hamiltonians with up to $50$ thousand terms indicate that the ratio of total terms over number of commuting groups scales ``at least linearly
with the number of qubits'' \cite{yenMeasuringAllCompatible2020}. Jena \emph{et al.} \cite{jenaPauliPartitioningRespect2019} also conjecture that this ratio is linear with respect to the lengths of the operators and provide numerical evidence for molecular Hamiltonians.

\section{Parallelisation} \label{sec:parallelisation}

In this section we will discuss our parallelisation of the \textsc{SELECT} stage of LCU, as the parallelisation of the \textsc{PREPARE} stage will significantly depend on the particular state preparation method chosen, but we note that a reduction in depth by a factor of $O(n)$ with $O(n \log n)$ qubits using fanout is achievable \cite{yuanOptimalControlledQuantum2023,zhangQuantumStatePreparation2022a}. In the usual application of \textsc{SELECT} (\cref{fig:SELECT}) the unitaries are applied in sequence, as is required by the complex controls on the ancilla register storing the coefficients. By fanning out the ancilla register, we can very straightforwardly remove this obstacle, and are then free to apply unitaries that act on disjoint sets of qubits in parallel. \\
To extend this to unitaries that are not already acting on disjoint sets of qubits, we can perform a unitary $U$ to transform the action of the unitaries on the system register into action on disjoint qubits. The set of unitaries must satisfy two conditions in order for this to be possible:

\begin{itemize}
    \item They must all commute
    \item They must form a linearly independent set
\end{itemize}

We do this by first partitioning the terms of the Hamiltonian into commuting groups as described in \cref{sec:partitioning}, and then use a greedy search algorithm to find linearly independent sets within each of those commuting groups. \\
For a Hamiltonian $H$, we define the filling factor $\mathcal F$ as the average size of the sets of terms that satisfy these conditions, after placing all the terms into such sets, normalised by the maximum applicable size (which is the number of qubits in the main register).

\begin{figure}[H]
    \center
\begin{tikzpicture}[scale=1.3]
    \begin{yquant}[boxctrl/.style={/yquant/every
        control/.style={
            draw=black,
            fill=none,
            minimum size=0.1cm,
            inner sep=0pt,
            outer sep=0pt,
        }}]
        [name=ypos]
        [/yquant/register/minimum height=4mm] qubit {$\ket{\protect\ifcase\idx\relax \alpha \protect\else 0 \protect\fi}$} a[3];
        qubit {$\ket{\psi_\idx}$} q[2];
        qubit {$\ket{\psi_n}$} qn;
        
        slash a;

        cnot a[1-] | a[0];
        box {$U$} (q,qn);

        [name=left]
        
        [boxctrl] box {$P_0$} q[0] | a[0];
        [boxctrl] box {$P_1$} q[1] | a[1];
        [name=secondlastbox]
        hspace {4mm} -;
        [name=finalbox]
        [boxctrl] box {$P_n$} qn | a[2];

        [name=right]
        
        box {$U^\dagger$} (q,qn);

        cnot a[1-] | a[0];
    \end{yquant}
    \path (left |- ypos-1) -- (right |- ypos-2)
    node[midway,rotate=-45] {$\dots$};
    \node [rotate=-45] at ($ (finalbox) + (-0.5,0.35) $) {$\dots$};
    \node at ($ (ypos-1) + (0,-0.23) $) {$\dots$};
    \node at ($ (ypos-1) + (-0.15,-1.95) $) {$\dots$};
\end{tikzpicture}
\caption{Applying Pauli operators in the \textsc{SELECT} step of LCU in parallel, by using fanout and a Clifford transformation $U$. The square controls indicate multiple controls on the ancilla register given the index of the Pauli operator.}
\label{fig:SELECT_parallel}
\end{figure}
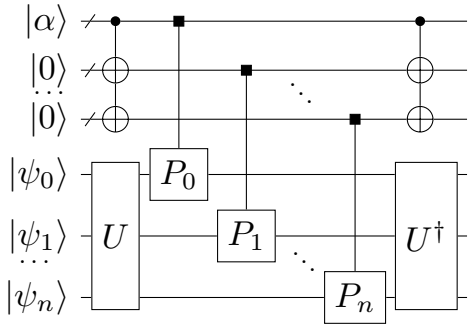

For the case where the unitaries in \textsc{SELECT} are Pauli operators $P_i$, we can partition the operators into linearly independent commuting sets, and then perform Clifford transformations to transform the terms in each set into single qubit Pauli operators as described in \cref{sec:partitioning}. We can then apply these controlled Pauli operators in parallel from the ancilla registers, applying up to $n$ terms simultaneously, \footnote{If the size of the commuting group is $m>n$, it can still be applied in one step by enlarging the system register to $m$ qubits by appending qubits in $\ket{0}$ and performing a Clifford circuit on all $m$ qubits (this also requires increasing the size of the fanout).} and then perform the inverse Clifford transformation, as depicted in \cref{fig:SELECT_parallel}.

The issue with this is that the linear depth saving from applying $n$ terms in parallel is offset by the depth of the Clifford transformations, which is $O(n)$ in the worst case when using reversible compilation \cite{maslovDepthOptimizationCZ2022}. We therefore present an alternative compilation of Clifford circuits using adaptive circuits and teleportation with $3n$ ancilla that is constant depth in \cref{sec:clifford_constant_depth}. \\

\begin{figure}[H]
    \includegraphics[width=\linewidth]{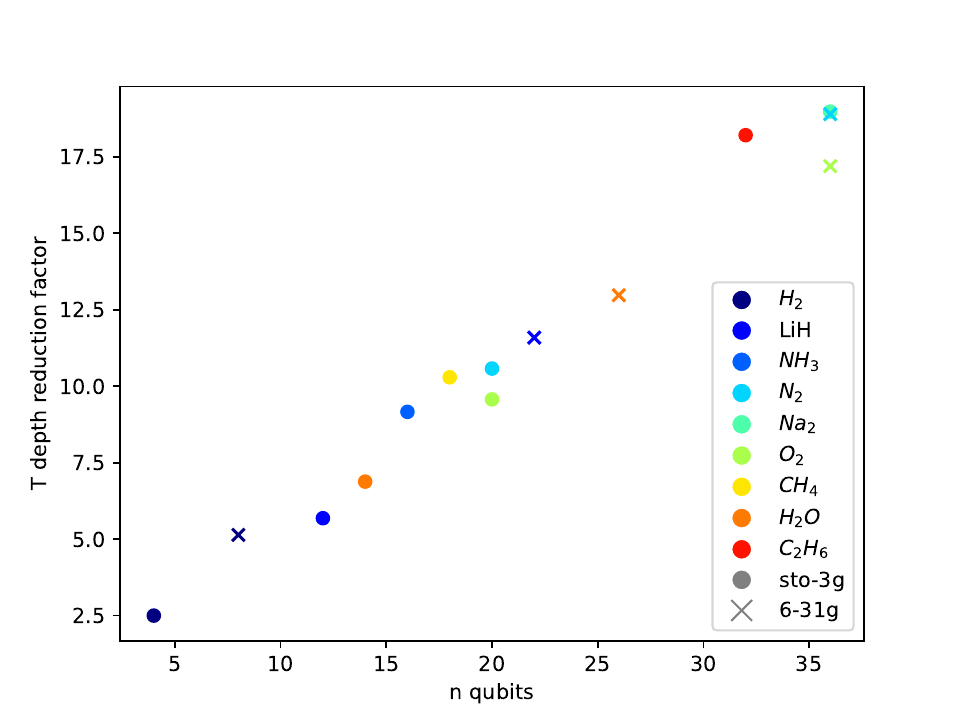}
    \caption{$T$ depth reduction as determined by the average size of the parallelisable sets found in molecular Hamiltonians up to $36$ qubits. The slope indicates an approximately constant $\mathcal F \approx 0.5$ for molecular Hamiltonians in this regime. Colour indicates the molecule and marker indicates the choice of basis.}
    \label{fig:molecular_depth_reduction}
\end{figure}

The depth reduction achieved is determined by the filling factor $\mathcal F$. We conjecture that this is $O(n)$ in the case of practical molecular Hamiltonians and provide evidence from the literature based on the size of commuting groups in Hamiltonians for this in \cref{sec:partitioning}. In \cref{fig:molecular_depth_reduction} we demonstrate the factor of improvement in $T$ depth for a series of molecular Hamiltonians by performing the partitioning into parallelisable sets, with the slope of the graph indicating an approximately constant $\mathcal F \approx 0.5$ for molecular Hamiltonians in this regime.
We also performed the partitioning for `Hamiltonians' containing only Pauli $X$ terms (as would be found in QROM circuits) up to $14$ qubits, and find $\mathcal F \approx 1$ for this case (see \cref{sec:qrom_filling}).

Given this partitioning into parallelisable sets, the space-time saving is a factor of $O\left( \frac{n\mathcal{F}}{\log n}\right)$. Although the saving will be very significant in the high-$n$ regime, we find that even in the low-qubit regime, we can still achieve large speed-ups and a reduction in the overall space-time volume. For example, the $26$-qubit Hamiltonian of H$_2$O in the $6$-$31$G basis contains $13884$ terms, which can be applied in $1070$ parallelised steps by using $468$ qubits, resulting in $13$ times fewer layers of multi-controlled Pauli gates for a qubit increase of $11.7$ times, which slightly reduces the space-time volume by a factor of $\sim1.1$. See \cref{sec:fault_tolerant} for further discussion on how the space-time volume of the parallelised subroutine compares to the serial version when taking into account the space required by $T$ factories.

\subsection{Performing Clifford Circuits in Constant Depth} \label{sec:clifford_constant_depth}

In order to reduce the depth of the required Clifford circuits, we can use techniques from \cite{zhengDepthReductionQuantum2018} that apply them in constant depth using a constant number of resource stabiliser states on $O(n)$ ancilla. 
This method is based on the observation that any Clifford circuit is equivalent to circuits that contain a sequence of stages containing the same kind of gates, for example, Aaronson and Gottesman provided an $11$ stage compilation H-C-P-C-P-C-H-P-C-P-C where -H-, -P-, and -C- stand for stages composed of only Hadamard, Phase, and CNOT gates, respectively \cite{aaronsonImprovedSimulationStabilizer2004}. More recently, techniques for implementing Clifford circuits using only 3 layers of two-qubit gates (the sequence CX-CZ-P-H-CZ-P-H) have been developed \cite{proctorSimpleAsymptoticallyOptimal2023}. \\

\begin{figure}[H]
    \centering
    \includegraphics[width=0.6\linewidth]{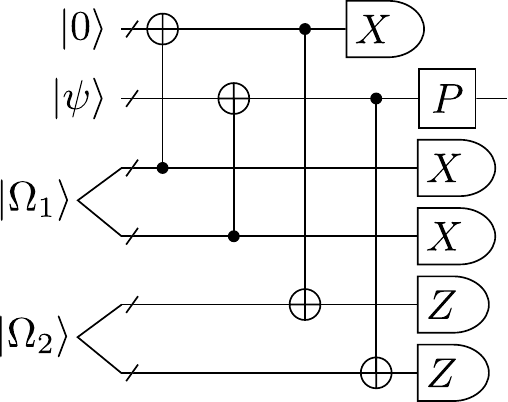}
    \caption{  
    Circuit based on Steane syndrome extraction for performing a Clifford $C$-stage from \cite{zhengDepthReductionQuantum2018}, using the CSS states $\ket{\Omega_{1,2}}$, with a Pauli correction $P$ depending on the measurement outcomes. This circuit uses $5n$ ancilla, but the same operation can be performed with $3n$ ancilla by performing the measurement on $\ket{\Omega_1}$ and then generating $\ket{\Omega_2}$ on the same ancilla.}
    \label{fig:teleportation}
\end{figure}

Considering only the non-trivial layers which contain 2-qubit gates, we can perform these layers by preparing states that are stabiliser states of Calderbeck-Shor-Steane (CSS) codes (up to single qubit rotations) \cite{chenMultipartiteQuantumCryptographic2008} on $3n$ ancilla and performing Steane syndrome extraction circuits using these resource states as an input \cite{zhengDepthReductionQuantum2018}, followed by a correction consisting of single qubit Pauli gates (\cref{fig:teleportation}).

These resource states can themselves be prepared in constant depth using adaptive circuits with the standard method to prepare stabiliser states of measuring the stabilisers and correcting incorrect outcomes \cite{nielsenQuantumComputationQuantum2010a}, the required measurements of parities can be performed in constant depth using either circuits that are log-depth in the weight of the stabilisers, or in constant depth by noting that parity circuits are equivalent to fanout under a unitary transformation \cite{takahashiCollapseHierarchyConstantDepth2012} and using results described in \cref{sec:parallelisation_techniques}. 

Therefore, the required Clifford transformations can be performed in constant depth, potentially allowing for a linear depth saving from parallelisation.

\subsection{Replacing the Ancilla Fanout}

It may be the case that we are not willing to pay the cost of fanning out the ancilla register $n$ times, either because we are restricted in the number of available qubits, or, for QROM, it may be the case that the input register is of comparable (or larger) size than the output register. We therefore present a scheme for parallelising the action of the input register using a constant factor increase in ancilla qubits and additional constant-depth Clifford transformations.

\begin{figure}[H]
    \center
    \begin{tikzpicture}
        \begin{yquant}
            qubit {} q[4];
            qubit {} a[4];
            
            cnot q[3] | a[1,2] ~ a[0,3];
            cnot q[2] | a[0,2] ~ a[1,3];
            cnot q[1] | a[1,3] ~ a[0,2];
            cnot q[0] | a[2,3] ~ a[0,1];
            
        \end{yquant}
    \end{tikzpicture}
    \caption{Example of a QROM circuit with a complex pattern of positive (filled) and negative (empty) controls, with the action on the output register already diagonalised.}
    \label{fig:multicontrols}
\end{figure}
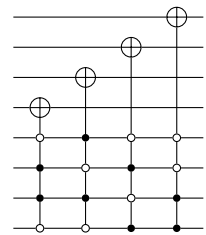

\begin{figure}[H]
    \center
    \begin{tikzpicture}
        \begin{yquant}
            qubit {} q[4];
            qubit {} a[4];
            box {$C$} (a);
            cnot q[3] | a[3] ~ a[0,1,2];
            cnot q[2] | a[2] ~ a[0,1,3];
            cnot q[1] | a[1] ~ a[0,2,3];
            cnot q[0] | a[0] ~ a[1,2,3];
            box {$C^\dagger$} (a);
        \end{yquant}
    \end{tikzpicture}
    \caption{Use of a Clifford transformation to `diagonalise' the action of the controls on the input register.}
    \label{fig:diagonalised_multicontrols}
\end{figure}
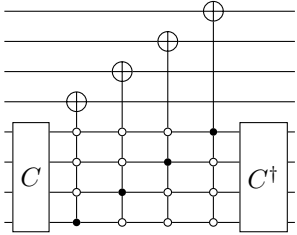

We can use a Clifford transformation on the ancilla register to transform the pattern of controls into the form in \cref{fig:diagonalised_multicontrols} (provided the bit strings describing the controls form a linearly independent set, see \cref{sec:qrom_filling} for a discussion on partitioning bit strings into linearly independent sets). This can be seen by the fact that conjugating the circuit in \cref{fig:multicontrols} by a CNOT gate acting between qubits $i$ and $j$ results in a row operation adding row $i$ of the controls (thought of as a binary matrix) onto row $j$, we can therefore use Gauss-Jordan elimination to diagonalise the pattern of controls provided the set of bit strings are linearly independent. The Clifford circuit we need to apply is also only a single $C$-stage (\cref{fig:teleportation}) so has a depth of $\sim 3\times$ less than a full Clifford unitary.

\begin{figure}[H]
    \center
    \begin{tikzpicture}
        \begin{yquant}[plusctrl/.style={/yquant/every
    control/.style={/yquant/operators/every not}, /yquant/every
    positive control/.style={}}]
            qubit {$\ket{\psi}$} q;
            qubit {} c;
            qubit {} b;
            qubit {} a;
            slash -; 
            hspace {2mm} -;
            box {$C$} (a);
            cnot b | a;
            box {$\text{Ham}$} a;
    
            [plusctrl] box {$1$} a | c;
    
            cnot q | b, c;
    
            [plusctrl] box {$1$} a | c;
            box {${\text{Ham}}^\dagger$} a;
            cnot b | a;
            box {$C^\dagger$} (a);
        \end{yquant}
    \end{tikzpicture}
    \caption{Circuit to compute and use the Hamming weight of the inputs to parallelise \cref{fig:diagonalised_multicontrols}.}
    \label{fig:hamming_weight_trick}
\end{figure}
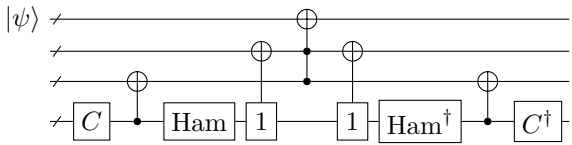

These controls cannot yet be trivially parallelised. However, each of the gates is now only activated from a state with Hamming weight 1, so by computing a flag containing \texttt{hamming\_weight == 1}, we can then perform the gates on the output register by performing a Toffoli on a register that contains a fanout of the original ancilla register, and a register containing copies of the \texttt{hamming\_weight==1} flag, as shown in \cref{fig:hamming_weight_trick}.

This Hamming weight circuit can be done in depth $\log a$, and a constant factor increase in ancilla qubits is required to perform this parallelisation.

\subsection{In fault-tolerant architectures} \label{sec:fault_tolerant}

Until this point, we have only been discussing the parallelisation implemented in terms of logical operations. However, there are some further considerations that must be made when examining the implications of parallelisation in fault-tolerant architectures. We will discuss these in the context of the surface code, but similar considerations apply to other fault-tolerant settings. \\

\begin{figure}[H]
    \includegraphics[width=\linewidth]{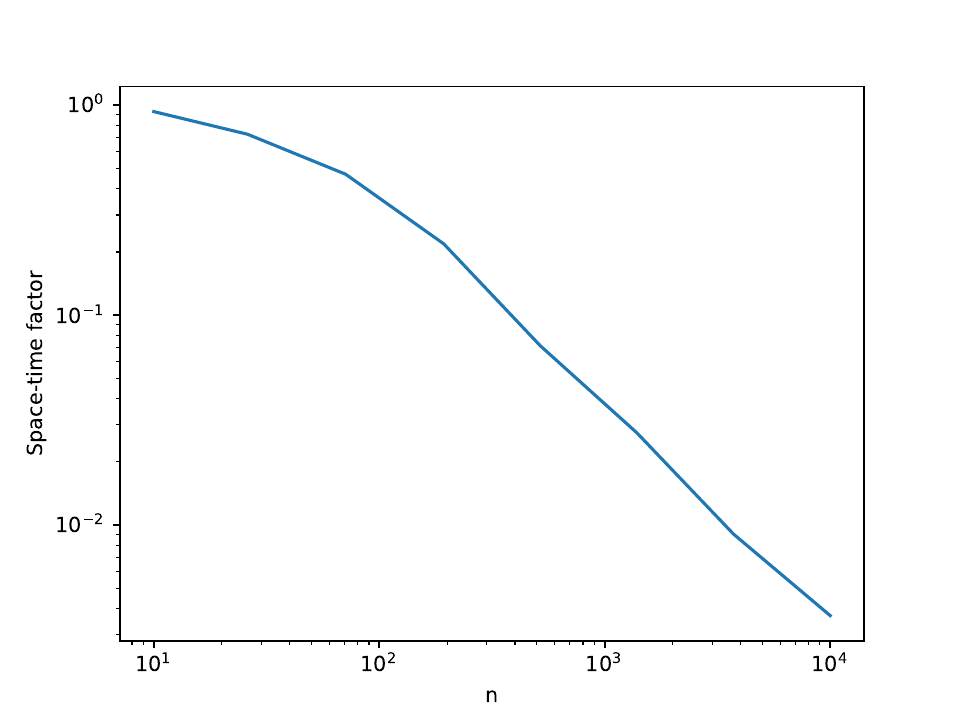}
    \caption{Scaling of the improvement in space-time volume of the parallelised algorithm over the serial version for $\mathcal F = 1$, when including the space required for additional $T$ factories due to the increased rate that the parallel circuit is able to use them.}
    \label{fig:spacetime_factor}
\end{figure}

In surface code architectures, the available operations are of the form Clifford$+T$ where the $T$ gates are non-transversal and require the use of additional techniques such as magic state distillation \cite{bravyiUniversalQuantumComputation2005a} to be implemented, and therefore have a higher cost than the Clifford gates, taking up the majority of the computational budget. Our LCU parallelisation succeeds in reducing the $T$-depth of the algorithm without changing the $T$-count. However, whereas only a factor $O(\log n)$ qubit increase is required in the logical setting, in the fault-tolerant setting, the number of qubits used for magic state distillation must be increased to keep up with the increased rate of magic state usage required for a speed-up for the parallelised algorithm. However, this additional cost in $T$ factories is modest and also introduces a $\log n$ factor (albeit with a higher constant) in the na\"ive case where every multi-controlled gate is done separately with $\log n$ $T$ cost. We also note that when the controls of the LCU/QROM circuits are constructed with unary iteration \cite{babbushEncodingElectronicSpectra2018}, which only results in a constant rate of $T$ state usage regardless of register size, the increase in space for the $T$ factories is further decreased.
In \cref{fig:spacetime_factor}, we use the Azure Quantum Resource Estimator \cite{beverlandAssessingRequirementsScale2022a} to compare the space-time volume of the parallel vs serial computation (assuming $\mathcal F = 1$) which initially has $n$ qubits in the main register, $\log n$ ancilla qubits, and use of unary iteration for the application of controls. We find orders of magnitude savings in space-time volume for large $n$.

The increase in qubits required for $T$ factories may also be mitigated by the following factors:

\begin{itemize}
    \item In cases where more magic state factories are required for other parts of the computation than are required to keep up with the rate of the serial \textsc{SELECT} procedure, the parallelised procedure is more efficiently able to use magic states at the rate they are produced.
    \item Magic states can be prepared offline and stored in a quantum memory.
    \item Magic state preparation has been significantly optimised in recent years \cite{litinskiMagicStateDistillation2019a,mokRigorousNoiseReduction2023}, and optimisation is likely to continue, further bringing down the cost.
\end{itemize}

\section{Discussion and Conclusion}

In this work we have produced an effective and low-overhead scheme for reducing the depth of the \textsc{SELECT} subroutine of LCU or data loading via QROM by a factor $O(n\mathcal{F})$, where $\mathcal F \le 1$ is the filling factor denoting the average size of the parallelisable sets of terms to be applied, whilst only increasing the required number of qubits by a factor $O(\log n)$. A key part of this scheme is a method for performing Clifford transformations in constant depth based on adaptive circuits and teleportation. Methods are also presented for the parallelisation of complex patterns of multi-controlled operations with only a constant factor qubit overhead. We provide a numerical study on the parallelisability of \textsc{SELECT} on molecular Hamiltonians up to $36$ qubits, and find they all can be reduced in depth by a factor of approximately $n/2$. We also note the procedure is inherently scalable with the number of ancillas available, meaning that if only $O(m \log n)$ ancillas are available for the fanout (as opposed to $O(n \log n)$), then the algorithm will simply perform $m$ operations in parallel. Alternatively, $m>n$ can be chosen, by adding $m-n$ qubits to the system register initialised in $\ket{0}$, and performing the Clifford transformation on all $m$ qubits. Indeed, depending on the distributions of the sizes of the commuting groups, it may not be worth using all $O(n \log n)$ ancillas if there are not a significant number of groups of size greater than $m$.

We believe that the use of parallelisation techniques in quantum algorithms is a fruitful direction for reducing the overheads of quantum computation, particularly when asymptotically optimal algorithms exist for problems such as Hamiltonian simulation, but the current runtime estimates for useful applications can be daunting. Works of this kind could also be of interest from the perspective of certain hardware platforms, as it means that scaling up hardware can reduce runtimes, allowing for offsetting of long gate times.

Further work includes accurate resource estimations of qubit counts and required wall-clock times in the fault-tolerant setting, the extension of the scheme to other families of unitaries, e.g. using matchgate circuits \cite{jozsaMatchgatesClassicalSimulation2008} and the generalisation of this scheme to alternative groupings of operators.

\subsection*{Note}
During the late stages of production of this manuscript, another pre-print making use of measurement-based circuits, teleportation and commuting groups of operators was produced \cite{kaldenbachMappingQuantumCircuits2023}. However, this work differs from ours in that it produces constant depth Clifford transformations in an alternative way inspired by measurement-based quantum computing, and specifically applies them to exponentials of Pauli operators found in VQE and QAOA.

\section*{Code Availability}
Code relevant to this paper, including python code for preprocessing Hamiltonians and Q\# code for the circuits described can be found \href{https://github.com/gboyd068/LCU_parallel}{here}. The code makes use of the packages OpenFermion \cite{mccleanOpenFermionElectronicStructure2020}, Stim \cite{gidney2021stim}, and graph-tool \cite{peixoto_graph-tool_2014}.

\section*{Acknowledgements}
I thank Simon Benjamin and Bálint Koczor for their advice and give thanks to them and Matthew Goh for comments on this manuscript. I would also like to thank Arthur Rattew, Tim Chan and Adam Connolly for helpful discussions.

\appendix

\section{Filling factor for QROM circuits} \label{sec:qrom_filling}

\begin{figure}[H]
    \includegraphics[width=\linewidth]{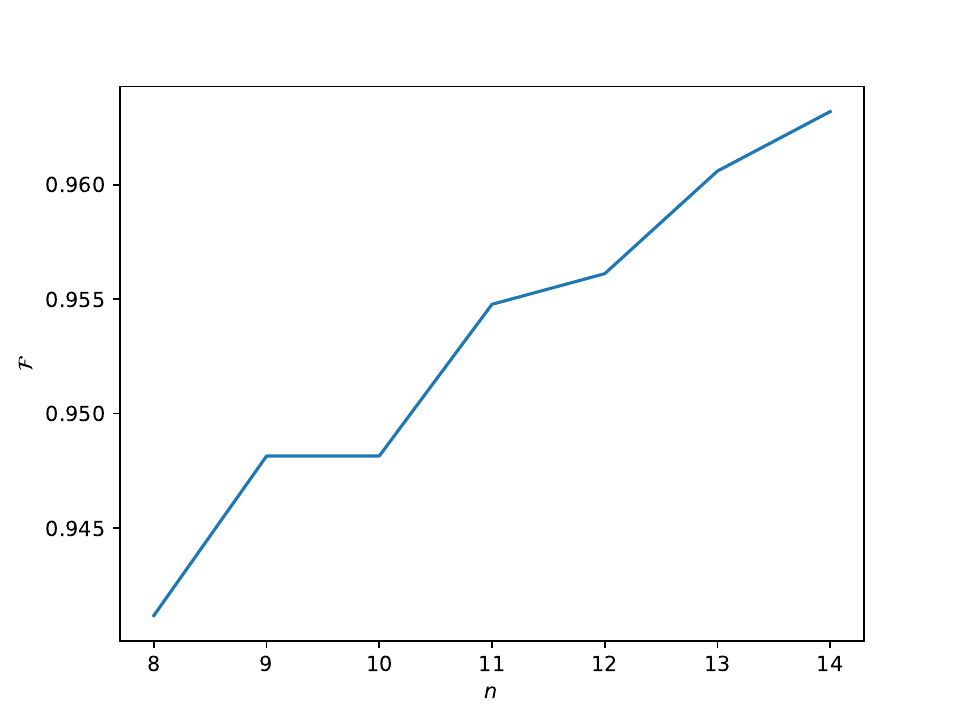}
    \caption{Filling factor for parallelisation of QROM circuits, using Hamiltonians containing all tensor products of Pauli $X$'s.}
    \label{fig:qrom_filling}
\end{figure}

For the case of parallelising QROM circuits, the operators being applied are all tensor products of Pauli $X$ operators, so already commute with each other. The linear independence condition becomes equivalent to the linear independence of binary vectors, which is easy to satisfy as demonstrated in \cref{fig:qrom_filling} where we plot $\mathcal{F}$ found by a greedy search over all $2^n$ bit strings for varying $n$.

\bibliographystyle{quantum}
\bibliography{refs}

\end{document}